\documentclass[11pt]{article}
\usepackage{epsf,amsmath,amssymb,axodraw,graphicx,scalefnt}
\usepackage{dsfont}
\usepackage{color}
\usepackage{rotating}

\usepackage[raiselinks=true,
bookmarks=true,
bookmarksopenlevel=1,
bookmarksopen=true,
bookmarksnumbered=true,
hyperindex=true,
plainpages=false,
pdfpagelabels=true,
pdfborder={0 0 0.5},
colorlinks=false,						
linkbordercolor={0 0.61 0.50},   
citebordercolor={0 0.61 0.50}]{hyperref}  %{0.57 0.74 0.57}

\allowdisplaybreaks

\newcommand{\abbrev}{\scalefont{.9}}

\newcommand{\drbar}{\overline{\mbox{\abbrev DR}}}
\newcommand{\drbarprime}{\overline{\mbox{\abbrev DR}}^\prime}

\title{
  \vskip-3cm{\baselineskip14pt
    \begin{flushleft}
      \normalsize SFB/CPP-11-27 \\
      \normalsize TTP11-16 \\
    \end{flushleft}}
  \vskip1.5cm
Three-loop anomalous dimensions for squarks in supersymmetric QCD
}
\author{Thomas Hermann, 
  Luminita Mihaila,
  Matthias Steinhauser\\[2em]
  {\it Institut f\"ur Theoretische Teilchenphysik}\\
  {\it Karlsruhe Institute of Technology (KIT), D-76128 Karlsruhe, Germany}\\
  }

\date{}%\fbox{RCS Version \fileversion{} --- \filedate{}}}

\begin{document}
\maketitle

%- {{{ abstract:

\begin{abstract}

In this paper we evaluate the renormalization constants and anomalous
dimensions for the squark wave function and mass within supersymmetric QCD. 
These results complement the
ones obtained in Ref.~\cite{Harlander:2009mn} and thus provide further
confirmation on the applicability of dimensional reduction to supersymmetric
QCD at three-loop order.
The three-loop anomalous dimension constitute important input to precision
predictions of the supersymmetric mass spectrum as obtained from the
evolution from the GUT to the TeV energy scale.

\end{abstract}

%- }}}

\thispagestyle{empty}

\newpage

%- {{{ body:

%- {{{ Introduction:

\section{Introduction}

Supersymmetry (SUSY) (for a review see, e.g., Ref.~\cite{Martin:1997ns})
has a number of appealing properties which classifies it as a
promising extension of the Standard Model (SM). Among them are the possibility
of gauge coupling unification, a dark matter candidate, and a solution to the
hierarchy problem. 

Although there is yet no clear evidence for the realization of SUSY in nature
it is mandatory to be prepared both on the experimental and theoretical
side. Currently there are several experimental groups who eagerly look for
signatures of supersymmetry in the data provided by the CERN Large Hadron
Collider (LHC). As far as theory is concerned it is on the one hand important
to provide precise predictions for production cross sections involving SUSY
particles. On the other hand there are a number of quantities which require
higher order loop corrections. A prominent example is the prediction of
the lightest Higgs boson mass
which recently became available to three
loops~\cite{Martin:2007pg,Harlander:2008ju,Kant:2010tf} resulting in an
uncertainty which can nevertheless be of the order of about
1~GeV~\cite{Kant:2010tf}. Another example where higher order corrections
within a supersymmetric theory are very welcome are the renormalization group
functions. They are crucial for the running from low to high energy scales and
constitute an important input for the spectrum generators (see, e.g.,
Refs.~\cite{Allanach:2001kg,Djouadi:2002ze,Porod:2003um}) which predict the
SUSY spectrum on the basis of only a few assumptions at energies of about
$10^{16}$~GeV.

The canonical choice for the regularization used for higher order loop
calculations is dimensional regularization (DREG). However, it is known since
about 30 years that DREG breaks SUSY. As a way out dimensional reduction
(DRED) has been
formulated~\cite{Siegel:1979wq,Siegel:1980qs,Stockinger:2005gx} which takes
over most of the convenient features from DREG and is thus a viable
alternative for practical multi-loop calculations. It is worth mentioning that
DRED is equivalent to DREG for non-SUSY theories as has been shown in
Refs.~\cite{Jack:1993ws,Capper:1979ns,Jack:1994bn,Harlander:2006rj,Harlander:2006xq,Jack:2007ni,Kilgore:2011ta}.
Furthermore it has been demonstrated in a number of
papers~\cite{Jack:1996qq,Ferreira:1996ug,Jack:1996vg,Jack:2004ch,Harlander:2009mn,Mihaila:2009bn}
that DRED is consistent with SUSY QCD at the three-loop level.  In this paper
we provide as new ingredients a further contribution by computing three-loop
renormalization constants for the mass and mixing angle of squarks
in the minimal subtraction scheme, which in the
context of DRED is called $\drbar$.

The renormalization constants and the corresponding anomalous dimensions up to
two-loop order has been computed in
Ref.~\cite{Barger:1993gh,Martin:1993zk,Jack:1994kd,Yamada:1994id,Kant:2010tf}.
Three-loop corrections have been considered in
Refs.~\cite{Jack:1996qq,Ferreira:1996ug,Jack:1996vg,Jack:2004ch,Kazakov:2000ih}
using relations between the beta functions of the gauge and Yukawa couplings
and the anomalous dimensions of the symmetry breaking parameters that can be
established in a softly broken supersymmetric
theory~\cite{Avdeev:1997vx,Jack:1997pa,Jack:1998iy}.  In
Ref.~\cite{Harlander:2009mn} the wave function renormalization constants of
quarks, squarks, gluons, gluinos, ghosts and $\epsilon$ scalars and the
renormalization constants for the quark and gluino mass were calculated to
three-loop order in the framework of SUSY QCD. In Ref.~\cite{Harlander:2009mn}
also the $\beta$ function for the strong coupling constant has been derived
from all possible three-point functions. The fact that in each case the same
expression has been obtained provides a check on the consistency of DRED with
gauge invariance and supersymmetry.    
In this paper the squark renormalization constants are computed to
three loops using the component field
approach.  The main difficulty of this calculation in contrast to the
renormalization constants for the gluino and quark masses is that
the squark mass renormalization constant depends on the masses of
the occurring particles in the loops although a renormalization scheme based
on minimal subtraction is adopted.  Furthermore, there is an interplay of the
renormalization of the $\epsilon$ scalar and the squark mass which will also
be discussed in this paper.

The remainder of the paper is organized as follows: In the next
Section we derive formulae for the squark renormalization constants
and briefly outline the procedure used for the construction of the
exact mass dependence.  Furthermore, the renormalization of the
$\epsilon$ scalars is discussed in detail.  Our results are presented
in Section~\ref{sec::res} and Section~\ref{sec::conclusions} contains
the conclusions.

%- }}}
%- {{{ Formalism:

\section{\label{sec::formalism}Formalism}

The calculations in this paper are performed in the framework of
SUSY QCD with $n_q=5$ massless quarks and a massive top quark ($m_t$).
The scalar super partners of the latter has two mass eigenstates 
($m_{\tilde{t}_1}$ and $m_{\tilde{t}_2}$) which may
have different masses and thus a non-vanishing mixing angle occurs.
The super partners of the $n_q$ light quarks are assumed to have degenerate
masses ($m_{\tilde{q}}$) and vanishing mixing angle. 
A generalization to a non-degenerate
spectrum is possible in a straightforward way from the formalism for the top
squark sector which is discussed in detail in the following.
The gluino mass is denoted by $m_{\tilde{g}}$.

Most of the formulae which we are going to present in the following
can already be found in Ref.~\cite{Kant:2010tf}. For
completeness we repeat the most important ones here and extend them to three
loops.
Unless stated otherwise all parameters in the following derivation are
$\drbar$ quantities which depend on the renormalization scale $\mu$. For the
sake of compactness the latter is omitted.
Bare quantities are marked by a superscript ``(0)''.

It is common to denote the left- and right-handed components of the top squark
by $\tilde{t}_L$ and $\tilde{t}_R$, respectively. The corresponding mass
matrix is given by  
\begin{align}
  {\cal M}^2_{\tilde{t}} &= 
  \left(
    \begin{array}{cc}
      m_t^2 + M_Z^2\big( \frac{1}{2} - \frac{2}{3} \sin^2\vartheta_W \big)
      \cos 2\beta  + M^2_{\tilde{Q}} & m_t \big( A_t - \mu_{\rm SUSY}
      \cot \beta \big) \\ 
      m_t \big( A_t - \mu_{\rm SUSY} \cot \beta \big) &  m_t^2 +
      \frac{2}{3} M^2_Z 
      \sin^2\vartheta_W \cos 2\beta  + M^2_{\tilde{U}} \\ 
    \end{array}
  \right) \nonumber \\  \nonumber \\
  &\equiv \left(
    \begin{array}{cc}
      m^2_{\tilde{t}_L} & m_t X_t \\
      m_t X_t  & m^2_{\tilde{t}_R} \\
    \end{array}
  \right)
  \label{eq::Mtil}
\end{align}
with $X_t = A_t - \mu_{\rm SUSY} \, \cot{\beta}$.
$A_t$ is the soft SUSY breaking tri-linear coupling, and $M_{\tilde{U}}$ and
$M_{\tilde{Q}}$ are the soft SUSY breaking masses.
With the help of the unitary transformation
\begin{align}
  \left(
    \begin{array}{c}
      \tilde{t}_1\\
      \tilde{t}_2\\
    \end{array}
  \right) = 
  R^{\dagger}_{\tilde{t}}
  \left(
    \begin{array}{c}
      \tilde{t}_L\\
      \tilde{t}_R\\
    \end{array}
  \right) \, ,
  \label{UnitTransSquark}
\end{align}
it is possible to diagonalize ${\cal M}^2_{\tilde{t}}$ 
\begin{align}
  \left(
    \begin{array}{cc}
      m^2_{\tilde{t}_1} & 0\\
      0 & m^2_{\tilde{t}_2}\\
    \end{array}
  \right) = R^{\dagger}_{\tilde{t}}\, {\cal M}^2_{\tilde{t}}\,
  R_{\tilde{t}} \,,
  \label{eq::Mtildiag}
\end{align}
where the eigenvalues are the masses of the eigenstates $\tilde{t}_1$ and
$\tilde{t}_2$. They read
\begin{align}
  m^2_{\tilde{t}_{1,2}} = \frac{1}{2}\Bigg[ m^2_{\tilde{t}_L} +
  m^2_{\tilde{t}_R} \mp \sqrt{\Big( m^2_{\tilde{t}_L} 
    - m^2_{\tilde{t}_R} \Big)^2 + 4 m_t^2 X_t^2} \Bigg] \,.
\end{align}
The unitary transformation can be parametrized by the mixing angle 
\begin{align}
  R_{\tilde{t}} =
  \left(
    \begin{array}{cc}
      \cos\theta_t & -\sin\theta_t\\
      \sin\theta_t & \cos\theta_t\\
    \end{array}
  \right)\,,
\end{align}
with
\begin{align}
  \sin\big(2\theta_t \big) = \frac{2 m_t \, \big(A_t - \mu_{\rm SUSY}
    \, \cot \beta 
    \big)}{m^2_{\tilde{t}_1} - m^2_{\tilde{t}_2}} \,. 
\end{align}

The renormalization constants connected to the top squark are extracted from the 
top squark propagator. At tree-level it is a diagonal
$2\times 2$ matrix which receives non-diagonal entries at loop-level. It is
convenient to absorb the corresponding counterterms into a renormalization
constant for the mixing angle which we introduce via
\begin{align}
  \theta^{(0)}_t = \theta_t + \delta\theta_t \label{dThetat}
  \,.
\end{align}
In order to be able to write
down the renormalized top squark propagator we define the
renormalization constants 
as follows: The wave function renormalization constant defined through
\begin{align}
  \left(
    \begin{array}{c}
      \tilde{t}_1^{\,(0)}\\
      \tilde{t}_2^{\,(0)}\\
    \end{array}
  \right) = 
  {\cal Z}^{1/2}_{\tilde{t}}
  \left(
    \begin{array}{c}
      \tilde{t}_1\\
      \tilde{t}_2\\
    \end{array}
  \right) 
\end{align}
can be parametrized by a universal factor $\tilde{Z}^{1/2}_2$
and the renormalization constant for the mixing angle
\begin{align}
 {\cal Z}^{1/2}_{\tilde{t}} = \tilde{Z}^{1/2}_2
 \left( \label{squarkWFMaxtrix}
 \begin{array}{cc}
  \cos\delta\theta_t & \sin\delta\theta_t \\
  -\sin\delta\theta_t & \cos\delta\theta_t\\
 \end{array}
 \right) \,.
\end{align}
This equation follows from Eq.~(\ref{UnitTransSquark}) and 
$(\tilde{t}_L^{(0)},\tilde{t}_R^{(0)})^T = \tilde{Z}^{1/2}_2 (\tilde{t}_L,\tilde{t}_R)^T$.
Furthermore, the renormalized mass matrix can be parametrized as follows
\begin{align}
  \left( %\label{squarkMassMatrix}
    \begin{array}{cc}
      (m^{(0)}_{\tilde{t}_1})^2 & 0 \\
      0 & (m^{(0)}_{\tilde{t}_2})^2 \\
    \end{array}
  \right) \rightarrow 
  \left(
    \begin{array}{cc}
      m^2_{11}Z_{m_{11}} & m^2_{12}Z_{m_{12}}\\
      m^2_{21}Z_{m_{21}} & m^2_{22}Z_{m_{22}}\\
    \end{array}
  \right) \equiv {\cal M} \,,
  \label{eq::Mt}
\end{align}
where we require that the off-diagonal elements in the renormalized mass matrix
vanish.
As a consequence, the counterterm $\delta\theta_t$ takes care of the
divergences in the self-energy contribution where a $\tilde{t}_1$ transforms
into a $\tilde{t}_2$ or vice versa. This can be seen in the explicit formulae
given below.
The diagonal elements of Eq.~(\ref{eq::Mt}) can be identified with the
renormalization of the masses 
\begin{equation}
  (m^{(0)}_{\tilde{t}_i})^2 
  = m_{ii}^2 Z_{m_{ii}}
  = m^2_{\tilde{t}_i} Z_{m_{\tilde{t}_i}} 
  \,.
\end{equation}

In order to formulate the renormalization conditions it
is convenient to consider the renormalized inverse top squark
propagator given by  
\begin{align}
  i{\cal S}^{-1}(p^2) = p^2\left({\cal Z}^{1/2}_{\tilde{t}}\right)^{\dagger}
  {\cal Z}^{1/2}_{\tilde{t}} 
  - \left({\cal Z}^{1/2}_{\tilde{t}}\right)^{\dagger}\left[ {\cal
      M} - \Sigma(p^2) \right] {\cal
    Z}^{1/2}_{\tilde{t}} 
  \label{squarkProp} 
\end{align}
where 
\begin{align}
  \Sigma(p^2) =
  \left(
    \begin{array}{cc}
      \Sigma_{11}(p^2) & \Sigma_{12}(p^2)\\
      \Sigma_{21}(p^2) & \Sigma_{22}(p^2)\\
    \end{array}
  \right) \,,
\end{align}
stands for the matrix of the squark self  energy.
In the $\drbar$ scheme the renormalization conditions read
\begin{equation}
  {\cal S}^{-1}_{ij}(p^2)\bigg|_{\rm pp} = 0 
  \,,
  \label{squarkRenoCondition}
\end{equation}
where ``pp'' stands for the ``pole part''.

In order to obtain explicit formulae for the evaluation of the
renormalization constants it is convenient to define perturbative expansions 
of the quantities entering Eq.~(\ref{squarkRenoCondition}). Up to three-loop
order we have
\begin{align}
  Z_k &= 1 + \left (\frac{\alpha_s}{4\pi} \right)\delta Z_k^{(1)} + \left
  (\frac{\alpha_s}{4\pi} \right)^2\delta Z_k^{(2)} 
  + \left (\frac{\alpha_s}{4\pi} \right)^3\delta Z_k^{(3)} + {\cal
    O}(\alpha_s^4) \nonumber \,,\\ 
  \delta\theta_t &= \left (\frac{\alpha_s}{4\pi} \right)\delta \theta_t^{(1)}
  + \left (\frac{\alpha_s}{4\pi} \right)^2\delta\theta_t^{(2)} 
  + \left (\frac{\alpha_s}{4\pi} \right)^3\delta \theta_t^{(3)} + {\cal
    O}(\alpha_s^4) \,,\nonumber\\ 
  \Sigma_{ij} &= \left (\frac{\alpha_s}{4\pi} \right) \Sigma_{ij}^{(1)} +
  \left (\frac{\alpha_s}{4\pi} \right)^2\Sigma_{ij}^{(2)} 
  + \left (\frac{\alpha_s}{4\pi} \right)^3\Sigma_{ij}^{(3)} + {\cal
    O}(\alpha_s^4)\,,
\end{align}
where $i,j\in\{1,2\}$ and $k\in\{2,m_{\tilde{t}_1},m_{\tilde{t}_2}\}$.
Inserting these equations into~(\ref{squarkProp}) one can solve 
Eq.~(\ref{squarkRenoCondition}) iteratively order-by-order in $\alpha_s$.
At one-loop order one gets
\begin{eqnarray}
  \bigg\{\Sigma_{ii}^{(1)} - m^2_{\tilde{t}_{i}}\left(\delta \tilde{Z}^{(1)}_2
  + \delta Z^{(1)}_{m_{\tilde{t}_i}} \right)  + p^2\delta
  \tilde{Z}^{(1)}_2\bigg\} \Bigg|_{\rm pp} &=& 0  
  \,, \quad i=1,2  \,,\nonumber\\
  \bigg\{ \Sigma_{12}^{(1)} - \delta\theta_t^{(1)} \left( m^2_{\tilde{t}_{1}} -
  m^2_{\tilde{t}_{2}} \right) \bigg\} \Bigg|_{\rm pp} &=& 0 \,. 
  \label{squarkRenBeding1L}
\end{eqnarray}

The terms proportional to $p^2$ in the first equation
of~(\ref{squarkRenBeding1L}) are used to compute the wave function
renormalization constant which is independent of all occurring masses. Thus
they can be set to zero and one obtains
\begin{equation}
  \delta \tilde{Z}^{(1)}_2 
  = -\frac{1}{p^2} \Sigma_{11}^{(1)}(p^2) \Bigg|_{\rm pp}
  = -\frac{1}{p^2} \Sigma_{22}^{(1)}(p^2) \Bigg|_{\rm pp}
  \,.
\end{equation}
Once $\delta \tilde{Z}^{(1)}_2$ is known Eq.~(\ref{squarkRenBeding1L}) is used
to obtain $\delta Z^{(1)}_{m_{\tilde{t}_i}}$ keeping the mass dependence in
$\Sigma_{ii}^{(1)}$ (see below for more details).
The second equation of~(\ref{squarkRenBeding1L}) is used to obtain the 
renormalization constant of the mixing angle via
\begin{equation}
 \delta \theta_t^{(1)} = \frac{\Sigma_{12}^{(1)}}{ m^2_{\tilde{t}_{1}} -
   m^2_{\tilde{t}_{2}}} \Bigg|_{\rm pp} \,. 
\end{equation}

Proceeding to two loops we obtain the equations
\begin{align}
  &\Bigg[ \Sigma_{ii}^{(2)} + \delta \tilde{Z}_2^{(1)} \Sigma_{ii}^{(1)} -
  m^2_{\tilde{t}_{i}} \Big( \delta \tilde{Z}_2^{(2)}  
  +  \delta \tilde{Z}_2^{(1)}\delta Z^{(1)}_{m_{\tilde{t}_i}} +  \delta
  Z^{(2)}_{m_{\tilde{t}_i}}\Big) 
  + \delta \tilde{Z}^{(2)}_2 p^2 
  \nonumber  \\
  &+ (-1)^{(i+1)} \delta \theta_t^{(1)} \Big( -2 \Sigma_{12}^{(1)} + \delta
  \theta_t^{(1)} \big( m^2_{\tilde{t}_{1}} - m^2_{\tilde{t}_{2}} \big) \Big) 
  \Bigg] \Bigg|_{\rm pp} = 0 \, , \quad i=1,2 \,, 
  \label{squarkRenBeding2L}
\end{align}
\begin{align}
  &\Bigg[
  -\delta \theta_t^{(2)} \Big( m^2_{\tilde{t}_{1}} - m^2_{\tilde{t}_{2}} \Big)
  - \delta \theta_t^{(1)} \delta \tilde{Z}_2^{(1)} \Big( m^2_{\tilde{t}_{1}} 
  - m^2_{\tilde{t}_{2}} \Big) - \delta \theta_t^{(1)} \delta
  Z^{(1)}_{m_{\tilde{t}_1}}  m^2_{\tilde{t}_{1}} 
  + \delta \theta_t^{(1)} \delta Z^{(1)}_{m_{\tilde{t}_2}}
  m^2_{\tilde{t}_{2}} 
  \nonumber \\
  &+ \delta \theta_t^{(1)} \Sigma_{11}^{(1)} - \delta \theta_t^{(1)}
  \Sigma_{22}^{(1)} + \delta \tilde{Z}_2^{(1)} \Sigma_{12}^{(1)} +
  \Sigma_{12}^{(2)} 
  \Bigg] \Bigg|_{\rm pp} = 0 \,,
\end{align}
which are solved for $\tilde{Z}_2^{(2)}$, $\delta Z^{(2)}_{m_{\tilde{t}_i}}$
and $\delta \theta_t^{(2)}$ using the same strategy as at one-loop level. 

Similarly, at three-loop order we have
\begin{align}
  &\Bigg[ (-1)^{i+1} \, \bigg\{ \,  \Big(\delta \theta_t^{(1)}\Big)^2 \,
  \bigg( \delta \tilde{Z}_2^{(1)} \big( m^2_{\tilde{t}_{1}} -
  m^2_{\tilde{t}_{2}} \big) + 
  \delta Z^{(1)}_{m_{\tilde{t}_1}} m^2_{\tilde{t}_{1}} - \delta
  Z^{(1)}_{m_{\tilde{t}_2}} m^2_{\tilde{t}_{2}} - \Sigma_{11}^{(1)}
  +\Sigma_{22}^{(1)} \bigg) \nonumber \\ 
  & \quad \quad \quad + \delta \theta_t^{(1)} \bigg( 2\, \delta \theta_t^{(2)}
  \big( m^2_{\tilde{t}_{1}} - m^2_{\tilde{t}_{2}} \big)  
  - 2 \delta \tilde{Z}_2^{(1)} \Sigma_{12}^{(1)} - 2 \Sigma_{12}^{(2)} \bigg) 
  -2 \delta \theta_t^{(2)} \Sigma_{12}^{(1)} \bigg\}\nonumber \\
  &+\delta \tilde{Z}_2^{(1)} \bigg( \Sigma_{ii}^{(2)} - \delta
  Z^{(2)}_{m_{\tilde{t}_i}} m^2_{\tilde{t}_{i}} \bigg)  
  - \delta \tilde{Z}_2^{(2)} \delta Z^{(1)}_{m_{\tilde{t}_i}}
  m^2_{\tilde{t}_{i}} + \delta \tilde{Z}_2^{(2)} \Sigma_{ii}^{(1)} -  
  \delta \tilde{Z}_2^{(3)} m^2_{\tilde{t}_{i}} + \delta \tilde{Z}_2^{(3)} p^2
  \nonumber \\ 
  & - \delta Z^{(3)}_{m_{\tilde{t}_i}} m^2_{\tilde{t}_{i}} + \Sigma_{ii}^{(3)} 
  \Bigg] \Bigg|_{\rm pp} = 0 \, , \quad i=1,2 \, ,
  \label{eq::ren3l_1}
\end{align}
\begin{align}
  &\Bigg[
  \,  \delta \theta_t^{(1)} \, \bigg(- \delta \tilde{Z}_2^{(1)} \delta
  Z^{(1)}_{m_{\tilde{t}_1}} m^2_{\tilde{t}_{1}}  
  + \delta \tilde{Z}_2^{(1)} \delta Z^{(1)}_{m_{\tilde{t}_2}} m^2_{\tilde{t}_{2}} 
  + \delta \tilde{Z}_2^{(1)} \Sigma_{11}^{(1)} - \delta \tilde{Z}_2^{(1)}
  \Sigma_{22}^{(1)} \nonumber \\ 
  & \quad \quad \quad - \delta \tilde{Z}_2^{(2)} \big( m^2_{\tilde{t}_{1}} -
  m^2_{\tilde{t}_{2}} \big)  
  - \delta Z^{(2)}_{m_{\tilde{t}_1}} m^2_{\tilde{t}_{1}} + \delta
  Z^{(2)}_{m_{\tilde{t}_2}} m^2_{\tilde{t}_{2}} 
  + \Sigma_{11}^{(2)} - \Sigma_{22}^{(2)} \bigg) \nonumber \\
  & + \delta \theta_t^{(2)} \bigg( - \delta \tilde{Z}_2^{(1)} \big(
  m^2_{\tilde{t}_{1}} - m^2_{\tilde{t}_{2}} \big)  
  - \delta Z^{(1)}_{m_{\tilde{t}_1}} m^2_{\tilde{t}_{1}} + \delta
  Z^{(1)}_{m_{\tilde{t}_2}} m^2_{\tilde{t}_{2}} + \Sigma_{11}^{(1)} -
  \Sigma_{22}^{(1)} \bigg) \nonumber \\ 
  &- \delta \theta_t^{(3)} \big( m^2_{\tilde{t}_{1}} - m^2_{\tilde{t}_{2}}
  \big)  
  + \delta \tilde{Z}_2^{(1)} \Sigma_{12}^{(2)}+ \delta \tilde{Z}_2^{(2)}
  \Sigma_{12}^{(1)} + \Sigma_{12}^{(3)} 
  +\frac{2}{3} \, \Big(\delta \theta_t^{(1)}\Big)^3 \, \big(
  m^2_{\tilde{t}_{1}} - m^2_{\tilde{t}_{2}} \big) \nonumber \\ 
  &-2 \Big(\delta \theta_t^{(1)}\Big)^2 \Sigma_{12}^{(1)}
  \Bigg] \Bigg|_{\rm pp} = 0 \,.
  \label{eq::ren3l_2}
\end{align}

\begin{figure}[t]
  \centering
  \includegraphics[width=\linewidth]{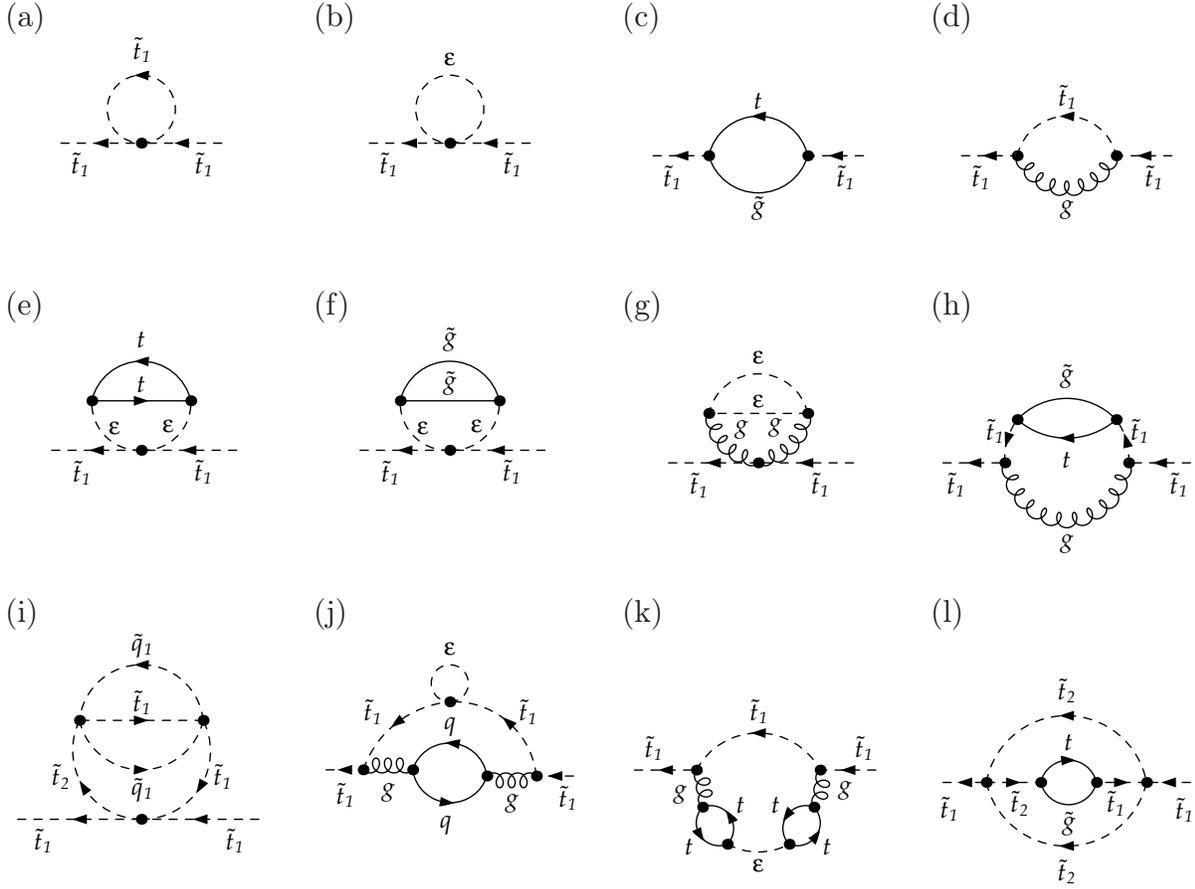}
  \caption[]{\label{fig::dias}Sample diagrams contributing to 
    $\Sigma_{11}$ at one, two and three loops.
    The symbols $t$, $\tilde{t}_i$, $g$, $\tilde{g}$ and
    $\epsilon$ denote top quarks, top squarks, gluons, gluinos,
    and $\epsilon$ scalars, respectively.
    }
\end{figure}

Sample diagrams contributing to $\Sigma_{11}$ up to three loops can be found
in Fig.~\ref{fig::dias}; the contributions to $\Sigma_{12}$ and $\Sigma_{22}$
look very similar.  Once the quantities $\Sigma_{11}$, $\Sigma_{12}$ and
$\Sigma_{22}$ are known to three-loop order it is possible to extract the
renormalization constants for the squark wave function and mass and the mixing
angle from Eqs.~(\ref{eq::ren3l_1}) and~(\ref{eq::ren3l_2}).

As compared to the corresponding self-energy contributions for fermions or
gauge bosons, which after proper projection only lead to logarithmically
divergent integrals, the quantities in the above equations have mass dimension
two. As a consequence the renormalization constants of the squark masses and
the mixing angles depend on the occurring masses, even in a minimal
subtraction scheme like $\drbar$.  At three-loop order an exact evaluation of
the corresponding integrals is not possible. It is nevertheless possible to
reconstruct the complete dependence on the occurring masses using repeated
asymptotic expansions and in addition some knowledge about the structure of
the final result. The latter can be induced from the known results at one- and
two-loop order. Besides the polynomial dependence inverse powers of first
(second) order in $m^2_{\tilde{t}_1} - m^2_{\tilde{t}_2}$ occur in the
two-loop contributions to $Z_{m_{\tilde{t}_i}}$ ($\delta\theta_t$). Thus we
expect that in $\delta Z_{m_{\tilde{t}_i}}^{(3)}$ at most
$1/(m^2_{\tilde{t}_1} - m^2_{\tilde{t}_2})^2$ and in $\delta\theta_t^{(3)}$ at
most $1/(m^2_{\tilde{t}_1} - m^2_{\tilde{t}_2})^3$ terms appear.  Asymptotic
expansion leads to results where these denominators are expanded in a
geometric series. If sufficient terms are evaluated it is straightforward to
properly reconstruct the inverse mass differences.

Using asymptotic expansion for several different hierarchies it is possible to
check that the final result is independent of the actual choice. In our
calculation we have chosen the external momentum as the largest scale in order
to avoid infrared divergences\footnote{Note that there are still massless gluons
  and light quarks in the theory.} and the $\epsilon$-scalar mass as the
smallest. As far as the squark masses, the gluino and the top quark mass is
concerned any hierarchy can be chosen. We decided to consider the three
choices
\begin{eqnarray}
  &&q^2 \gg m^2_{\tilde{t}_2} \gg m^2_{\tilde{q}} \gg m^2_{\tilde{t}_1}
  \gg m^2_{\tilde{g}} \gg m^2_t \gg m^2_{\epsilon}\,,\nonumber\\  
  &&q^2 \gg m^2_{\tilde{g}} \gg m^2_{\tilde{q}} \gg m^2_{\tilde{t}_2}  \gg
  m^2_{\tilde{t}_1} \gg  m^2_t \gg m^2_{\epsilon}\,,\nonumber\\  
  &&q^2 \gg m^2_{\tilde{g}} \gg m^2_{\tilde{q}} \gg m^2_{\tilde{t}_2}  \gg
  m^2_t \gg  m^2_{\tilde{t}_1} \gg m^2_{\epsilon}\,.
  \label{eq::hier}
\end{eqnarray}
We have checked that in all cases we obtain the same results for
$Z_{m_{\tilde{t}_i}}$  and $\delta\theta_t$.
Note that in the last hierarchy the top quark mass is even larger than the
corresponding squark mass which is allowed since the mass dependence in the
$\drbar$ counterterms has no physical meaning.

In each hierarchy of Eq.~(\ref{eq::hier}) six mass ratios appear. Some of the
expansions are simple and can be truncated after a few terms. E.g., all terms
with inverse contributions in $q^2$ can immediately be set to zero.
Similarly, all mass ratios where one has a top squark mass in the denominator
and $m_t$, $m_{\tilde{g}}$ or $m_{\tilde{q}}$ in the numerator only low-order
expansion terms appear in the final result. This has been checked by
increasing the expansion depth and verifying that the higher order terms are
zero.  Due to the occurrence of $1/(m^2_{\tilde{t}_1} - m^2_{\tilde{t}_2})$
terms in the exact result several terms in $m_{\tilde{t}_1}/m_{\tilde{t}_2}$
have to be kept in the expressions for the self energies in order to be able to
reconstruct the geometric series. In practice we compute terms up to
$(m_{\tilde{t}_1}/m_{\tilde{t}_2})^8$ and check that after including two more
powers in the top squark mass ratio the final result does not change.

At this point some comments on the treatment of the $\epsilon$ scalar mass,
$m_{\epsilon}$, are in order. In practice there are two
renormalization schemes for $m_{\epsilon}$ which are frequently used, the
$\drbar$ and on-shell scheme. In the latter one requires that the renormalized
mass vanishes in each order in perturbation theory whereas in the $\drbar$
prescription only the pole parts are subtracted by the renormalization
constant. We will present our results in a first step for $\drbar$ $\epsilon$
scalar masses and afterwards discuss the difference to the on-shell scheme.

In the $\drbar$ scheme it is important to keep $m_{\epsilon}$ different from
zero since the renormalization group equations for the squark masses and
$m_{\epsilon}$ are coupled. A non-vanishing $\epsilon$-scalar
mass in intermediate steps is also required for the computation of the 
anomalous dimensions in the $\drbarprime$ scheme~\cite{Jack:1994rk} (see
below) which was constructed in order to disentangle the running of
$m_{\epsilon}$ from the one of the squark parameters.

After the calculation of the bare self energies we renormalize all
occurring parameters in the $\drbar$ scheme. For our three-loop
calculation we need the counterterms for 
$\alpha_s$, $m_t$, $m_{\tilde{g}}$, $m_{\tilde{t}_i}$, $\theta_t$ and
$m_{\epsilon}$ to two-loop order and the one for $m_{\tilde{q}}$ to
one-loop approximation. Furthermore, also the QCD gauge parameter
has to be renormalized to two loops since it appears in the results for the
wave function anomalous dimensions.
All relevant counterterms can be found in the {\tt Mathematica} file
provided together with Ref.~\cite{Harlander:2009mn} and in
Ref.~\cite{Kant:2010tf}. The two-loop corrections for the
$\epsilon$-scalar mass renormalization is provided in Ref.~\cite{progdata}. 

For the calculation of the three-loop integrals we make use of several
computer programs which work hand-in-hand in order to reduce the
manual interaction to a minimum.  All Feynman diagrams are generated with
the program {\tt qgraf}~\cite{Nogueira:1991ex}.  The generated files
are manipulated by a {\tt perl} program~\cite{Harlander:2009mn}, which
implements the prescriptions of
Ref.~\cite{Denner:1992vza}, in order to obtain the
correct prefactors due to the Majorana character of the gluino.
Afterwards the output is transformed to {\tt
  FORM}~\cite{Vermaseren:2000nd} notation with the help of {\tt q2e}
and {\tt exp}~\cite{Harlander:1997zb,Seidensticker:1999bb}.  {\tt exp}
furthermore applies the asymptotic expansion (see, e.g.,
Ref.~\cite{Smirnov:2002pj}) in the hierarchies specified in
Eq.~(\ref{eq::hier}). As a result only one-scale integrals up to three
loops appear which can be evaluated with the packages {\tt
  MINCER}~\cite{Larin:1991fz} and {\tt
  MATAD}~\cite{Steinhauser:2000ry}.  Let us mention that we
implemented the DRED Feynman rules for SUSY QCD as given in
Ref.~\cite{Harlander:2004tp,Harlander:2009mn}.

Once the renormalization constants are available we compute the corresponding
anomalous dimension with the help of
\begin{equation}
  \gamma_X = \frac{\mu^2}{X} \frac{{\rm d}X}{{\rm d}\mu^2}\,,
  \label{eq::gam}
\end{equation}
where the quantity $X$ is either a mass parameter or the mixing angle
\begin{equation}
  X \in \bigg\{ m^2_{\tilde{t}_1},\, m^2_{\tilde{t}_2},\, m^2_{\tilde{q}},\,
  m_{\tilde{g}},\, m_t,\, m^2_{\epsilon},\, \theta_t \bigg\} \,. 
  \label{eq::X}
\end{equation}
In practice the derivation in Eq.~(\ref{eq::gam}) is taken after exploiting the
relation between the bare and the renormalized quantity. Since bare parameters
do not depend on $\mu$ the derivative acts only on the renormalization
constant. In the case of the top quark and the gluino the latter are mass
independent and thus the derivative w.r.t. $\mu$ can be rewritten into a
derivative w.r.t. $\alpha_s$. For the other parameters, however, one has to
take into account the mass dependence of the $Z$ factors. Let us as an example
consider the anomalous dimension of $m_{\tilde{t}_i}$ which leads to the
following chain of equations\footnote{In the subscript for the anomalous
  dimensions we write
  $m_{\tilde{t}_1}$ instead of $m^2_{\tilde{t}_1}$, etc..}
\begin{align}
  \gamma_{m_{\tilde{t}_i}} 
  &= - \frac{\mu^2}{Z_{m_{\tilde{t}_i}}} \, \frac{d}{d\mu^2} \,
  Z_{m_{\tilde{t}_i}}  
  \nonumber\\
  &= -\frac{\mu^2}{Z_{m_{\tilde{t}_i}}} \Bigg[
    \frac{dZ_{m_{\tilde{t}_i}}}{d\alpha_s} \, \frac{d\alpha_s}{d\mu^2}  
    + \sum_X \frac{dZ_{m_{\tilde{t}_i}}}{dX}\, \frac{dX}{d\mu^2} \Bigg] 
  \nonumber\\ 
  &= - \bigg[ \pi\, \beta \, \frac{d}{d\alpha_s} \left( \log Z_{m_{\tilde{t}_i}} \right)
    + \sum_X X\, \gamma_X \, \frac{d}{dX} \left( \log Z_{m_{\tilde{t}_i}}
    \right)\bigg]\,,
\end{align}
where $\beta(\alpha_s)$ is the anomalous dimension of the strong coupling and
$X$ runs over the parameters listed in Eq.~(\ref{eq::X}).

In the next Section we provide results for various anomalous dimensions. For
this purpose it is convenient to introduce the following expansion
\begin{eqnarray}
  \gamma_X &=& - \frac{\alpha_s}{\pi} \sum_{n\ge0} \left(
  \frac{\alpha_s}{\pi}\right)^n 
  \gamma_X^{(n)}\,. 
\end{eqnarray}

%- }}}
%- {{{ Results:

\section{\label{sec::res}Results}

In a first step we have computed
the three-loop corrections to the squark wave function
renormalization constant $\tilde{Z}_2$ (which is mass independent).
In the following we present results for the anomalous dimensions 
$\gamma_{m_{\tilde{t}_1}}$, $\gamma_{m_{\tilde{t}_2}}$,
$\gamma_{m_{\tilde{q}}}$ and $\gamma_{\theta_t}$
up to three-loop order which all have a non-trivial mass dependence.
The corresponding results for the renormalization constants can be found in
{\tt Mathematica} format on the internet page~\cite{progdata}.

At one-loop order we obtain the following results
\begin{align}
  m_{{\tilde t}_1}^2 \gamma_{m_{\tilde{t}_1}}^{(0)}
  &=
  C_F\left[ m_{\tilde g}^2 +
    \frac{1}{8}\big(1-c_{4t}\big)\, \left(  m_{{\tilde t}_1}^2 -  
    m_{{\tilde t}_2}^2 \right)  + m_{t}^2 - m_{\tilde g} \, m_{t} \,
    s_{2t} \right] \,,
  \label{eq::gamma0_a}
  \\
  \theta_t \gamma_{\theta_t}^{(0)} 
  &= 
  C_F c_{2t}
  \left[ -\frac{ m_{\tilde g}\, m_{t}}{m_{{\tilde t}_1}^2 -
      m_{{\tilde t}_2}^2}+\frac{s_{2t}}{4} 
    \right]\,,
  \label{eq::gamma0}
\end{align}
where the abbreviations 
$c_{nt} = \cos(n\theta_t)$ and $s_{nt} = \sin(n\theta_t)$ have been
introduced and $C_F=(N_C^2-1)/(2N_C)$ is the Casimir operator of the
fundamental representation of $SU(N_C)$.
In Eq.~(\ref{eq::gamma0_a}) we have given the result for
$\gamma_{m_{\tilde{t}_1}}$. The one for $\gamma_{m_{\tilde{t}_2}}$ is
obtained by interchanging $m_{\tilde{t}_1}$ and $m_{\tilde{t}_2}$ and
replacing $\theta_t$ by $-\theta_t$.

The two-loop coefficients read
\begin{align}
  m_{{\tilde t}_1}^2 \gamma_{m_{\tilde{t}_1}}^{(1)}
  &=
  C_A\, C_F\, \Bigg\{ \frac{3}{4}\, m_{\epsilon}^2 + \frac{11}{4} \, m_{\tilde g}^2 + \frac{3}{32} \, \left(1 - c_{4t} \right)\, \left( m_{{\tilde t}_1}^2 - m_{{\tilde t}_2}^2 \right)
+\frac{3}{4}\, m_{t}^2 - \frac{3}{2}\, m_{\tilde g}\, m_{t}\, s_{2t} \Bigg\} \nonumber \\
& +C_F^2\, \Bigg\{ - \frac{3}{2}\, m_{\tilde g}^2 -\frac{1}{16} \left(1 - c_{4t} \right) \, \left( m_{{\tilde t}_1}^2 - m_{{\tilde t}_2}^2 \right) 
-\frac{1}{2}\, m_{t}^2 + m_{\tilde g}\, m_{t}\, s_{2t} \Bigg\} \nonumber\\
&- C_F\, T_f\, \Bigg\{ n_q\, \Bigg[ \frac{1}{2}\, m_{\epsilon}^2 + \frac{3}{2}\, m_{\tilde g}^2 + m_{\tilde{q}}^2
 +\frac{1}{16}\, \left(1 - c_{4t} \right)\, \left( m_{{\tilde t}_1}^2 - m_{{\tilde t}_2}^2 \right) +\frac{1}{2}\, m_{t}^2 - m_{\tilde g}\, m_{t}\, s_{2t} \Bigg] \nonumber \\
&+n_t\, \Bigg[ \frac{1}{2}\, m_{\epsilon}^2 
+\frac{3}{2}\, m_{\tilde g}^2 +\frac{1}{16}\, \left( 9 -  c_{4t} \right) \, m_{{\tilde t}_1}^2 + \frac{1}{16}\, \left(7 + c_{4t} \right)\, m_{{\tilde t}_2}^2
- \frac{1}{2} \, m_{t}^2 -m_{\tilde g}\, m_{t}\, s_{2t} \Bigg] \Bigg\}\,,% \nonumber
\end{align}
\begin{align}
 \theta_t \gamma_{\theta_t}^{(1)} 
  &= C_F\, T_f\, \Bigg\{ \, n_q\, \Bigg[ \frac{m_{\tilde g}\, m_{t}}{ m_{{\tilde t}_1}^2 - m_{{\tilde t}_2}^2 } \, c_{2t} - \frac{1}{8}\, c_{2t}\, s_{2t} \Bigg]
 +n_t\, \Bigg[ \frac{m_{\tilde g}\, m_{t}}{m_{{\tilde t}_1}^2 - m_{{\tilde t}_2}^2} \, c_{2t} -\frac{1}{8} \, c_{2t}\, s_{2t} \Bigg] \Bigg\} \nonumber \\
& +C_A\, C_F\, \Bigg\{ \frac{3}{16}\, c_{2t}\, s_{2t} - \frac{3}{2}\, \frac{m_{\tilde g}\, m_{t}}{m_{{\tilde t}_1}^2 - m_{{\tilde t}_2}^2} \, c_{2t} \Bigg\}
+C_F^2\, \Bigg\{ \frac{m_{\tilde g}\, m_{t}}{m_{{\tilde t}_1}^2 - m_{{\tilde t}_2}^2}\, c_{2t} -\frac{1}{8}\, c_{2t}\, s_{2t} \Bigg\}\,,% \nonumber
\label{eq::theta2}
\end{align}
where $C_A$ is the Casimir operator of the adjoint representation of
$SU(N_C)$ and $T_F=1/2$ the trace normalization. $n_t$ counts the
top squark flavours and $n_q$ counts the
mass-degenerate squark flavours and at the same time the
massless quarks. In practice we have $n_t=1$ and $n_q=5$, however, it
is nevertheless convenient to keep the labels arbitrary.

Let us now come to the three-loop results.
The anomalous dimensions for the top squark masses are given by
\begin{align}
 m_{{\tilde t}_1}^2 \gamma_{m_{\tilde{t}_1}}^{(2)}
  &= C_F^3\, \Bigg\{ 3\, m_{\tilde g}^2 + \frac{1}{2} \, m_{t}^2 - \frac{3}{2} \, m_{\tilde g}\, m_{t}\, s_{2t}
 + \frac{1}{16} \, \left(1 -c_{4t} \right)\, \left( m_{{\tilde t}_1}^2 - m_{{\tilde t}_2}^2 \right) \Bigg\} \nonumber \\
&+C_A^2\, C_F\, \Bigg\{ \frac{45}{32} \, m_{\epsilon}^2 + \frac{15}{4} \, m_{\tilde g}^2 + \frac{3}{8} \, m_{t}^2 -\frac{9}{8} \, m_{\tilde g}\, m_{t}\, s_{2t}
 + \frac{3}{64}\,  \left( 1 - c_{4t} \right) \, \left( m_{{\tilde t}_1}^2 - m_{{\tilde t}_2}^2 \right) \Bigg\} \nonumber \\
& +C_F^2\, C_A\, \Bigg\{ - \frac{9}{16} \, m_{\epsilon}^2 - \frac{21}{8} \, m_{\tilde g}^2 - \frac{3}{8} \, m_{t}^2 + \frac{9}{8} \, m_{\tilde g}\, m_{t}\, s_{2t}
 - \frac{3}{64} \, \left( 1 - c_{4t} \right)\, \left( m_{{\tilde t}_1}^2 - m_{{\tilde t}_2}^2 \right) \Bigg\} \nonumber \\
&+C_F\, T_f^2\, \Bigg\{ n_t^2 \, \Bigg[ \frac{3}{8} \, m_{\epsilon}^2 - \frac{3}{2} \, m_{\tilde g}^2 + \frac{3}{4}\, m_{{\tilde t}_1}^2 
 -m_{t}^2 + \frac{3}{4} \, m_{\tilde g}\, m_{t}\, s_{2t} - \frac{1}{32} \left( 13  - c_{4t} \right)\, \left( m_{{\tilde t}_1}^2 - m_{{\tilde t}_2}^2 \right) \Bigg] \nonumber \\
& +n_q^2\, \Bigg[ \frac{3}{8} \, m_{\epsilon}^2 - \frac{3}{2} \, m_{\tilde g}^2 + \frac{3}{4} \, m_{\tilde{q}}^2 - \frac{1}{4} \, m_{t}^2
+ \frac{3}{4}\, m_{\tilde g}\, m_{t}\, s_{2t} -\frac{1}{32} \left(1 - c_{4t} \right)\, \left( m_{{\tilde t}_1}^2 - m_{{\tilde t}_2}^2 \right) \Bigg] \nonumber \\
&+n_q\, n_t\, \Bigg[ \frac{3}{4} \, m_{\epsilon}^2 - 3\, m_{\tilde g}^2 + \frac{3}{4} \, m_{\tilde{q}}^2 + \frac{3}{4}\, m_{{\tilde t}_1}^2
-\frac{5}{4}\, m_{t}^2 + \frac{3}{2}\, m_{\tilde g}\, m_{t}\, s_{2t} \nonumber \\
& -\frac{1}{16} \left(7 - c_{4t}\right)\, \left( m_{{\tilde t}_1}^2 - m_{{\tilde t}_2}^2 \right) \Bigg] \Bigg\} \nonumber \\
&+C_F^2\, T_f\, \Bigg\{ n_t\, \Bigg[ \frac{3}{8}\, m_{\epsilon}^2 - \frac{27}{4}\, m_{\tilde g}^2 + \frac{3}{4}\, m_{{\tilde t}_1}^2
- \frac{7}{4}\, m_{t}^2 + 3\, m_{\tilde g}\, m_{t}\, s_{2t} + 9\, m_{\tilde g}^2\, \zeta_3 + \frac{3}{2}\, m_{t}^2\, \zeta_3 \nonumber \\
& -\frac{9}{2}\, m_{\tilde g}\, m_{t}\, s_{2t} \, \zeta_3 +\left( m_{{\tilde t}_1}^2 - m_{{\tilde t}_2}^2 \right)\, \left(  \frac{1}{8}\, c_{4t} 
-\frac{1}{2} + \frac{3}{16} \, \zeta_3 -\frac{3}{16} \, c_{4t} \, \zeta_3 \right) \Bigg] \nonumber \\
&+n_q\, \Bigg[\frac{3}{8} \, m_{\epsilon}^2 - \frac{27}{4} \, m_{\tilde g}^2 + \frac{3}{4} \, m_{\tilde{q}}^2 
 -m_{t}^2 + 3\, m_{\tilde g}\, m_{t}\, s_{2t} + 9\, m_{\tilde g}^2\, \zeta_3 + \frac{3}{2}\, m_{t}^2\, \zeta_3
 - \frac{9}{2}\, m_{\tilde g}\, m_{t}\, s_{2t}\, \zeta_3 \nonumber \\
& + \left( m_{{\tilde t}_1}^2 - m_{{\tilde t}_2}^2 \right) \, \left(  \frac{1}{8} \, c_{4t} 
-\frac{1}{8} +\frac{3}{16} \, \zeta_3 - \frac{3}{16} \, c_{4t}\, \zeta_3 \right) \Bigg] \Bigg\} \nonumber \\
& + C_A\, C_F\, T_f\, \Bigg\{ n_q\, \Bigg[  \frac{1}{8} \, m_{t}^2-\frac{3}{2}\, m_{\epsilon}^2 - \frac{15}{8}\, m_{\tilde{q}}^2 
-\frac{3}{8} \, m_{\tilde g}\, m_{t}\, s_{2t} - 9\, m_{\tilde g}^2\, \zeta_3 - \frac{3}{2}\, m_{t}^2\, \zeta_3  \nonumber \\
& + \frac{9}{2}\, m_{\tilde g}\, m_{t}\, s_{2t} \, \zeta_3
+  \left( m_{{\tilde t}_1}^2 - m_{{\tilde t}_2}^2 \right) \, \left( \frac{1}{64} - \frac{1}{64}\, c_{4t} - \frac{3}{16} \, \zeta_3 
+\frac{3}{16} \, c_{4t} \, \zeta_3 \right) \Bigg] \nonumber \\
& +n_t\, \Bigg[  2\, m_{t}^2 - \frac{3}{2}\, m_{\epsilon}^2 - \frac{15}{8}\, m_{{\tilde t}_1}^2 - \frac{3}{8}\, m_{\tilde g}\, m_{t}\, s_{2t}
-9\, m_{\tilde g}^2\, \zeta_3
-\frac{3}{2} \, m_{t}^2\, \zeta_3 + \frac{9}{2} \, m_{\tilde g}\, m_{t}\, s_{2t} \, \zeta_3 \nonumber \\
& + \left( m_{{\tilde t}_1}^2 - m_{{\tilde t}_2}^2 \right) \, \left( \frac{61}{64}
-\frac{1}{64} \, c_{4t} - \frac{3}{16}\, \zeta_3 + \frac{3}{16} \,
c_{4t} \, \zeta_3 \right) \Bigg] \Bigg\}\,, % \nonumber
\end{align}
where $\zeta_3$ is Riemann's zeta function with the value
$\zeta_3=1.2020569\ldots$. 
The three-loop expression for $\gamma_{\theta_t}$ reads
\begin{align}
 \theta_t \gamma_{\theta_t}^{(2)} 
  &= C_F\, T_f^2\, \Bigg\{ n_q^2\, \Bigg[ \frac{m_{\tilde g}\, m_{t}}{ m_{{\tilde t}_1}^2 - m_{{\tilde t}_2}^2 } \, \frac{3}{4}\, c_{2t}
 -\frac{1}{16} \, c_{2t}\, s_{2t} \Bigg] 
 + n_t^2\, \Bigg[ \frac{m_{\tilde g}\, m_{t}}{m_{{\tilde t}_1}^2 - m_{{\tilde t}_2}^2} \, \frac{3}{4}\, c_{2t} - \frac{1}{16} \, c_{2t}\, s_{2t} \Bigg] \nonumber \\
& +n_q\, n_t\, \Bigg[ \frac{m_{\tilde g}\, m_{t}}{ m_{{\tilde t}_1}^2 - m_{{\tilde t}_2}^2 }\, \frac{3}{2} \, c_{2t} -\frac{1}{8} \, c_{2t} \, s_{2t} \Bigg] \Bigg\} 
\nonumber \\
 &+ C_F^3 \, \Bigg\{ \frac{1}{8}\, c_{2t}\, s_{2t} - \frac{m_{\tilde g}\, m_{t}}{m_{{\tilde t}_1}^2 - m_{{\tilde t}_2}^2} \, \frac{3}{2} \, c_{2t} \Bigg\} 
 + C_A^2\, C_F\, \Bigg\{ \frac{3}{32}\, c_{2t}\, s_{2t} - \frac{m_{\tilde g}\, m_{t}}{m_{{\tilde t}_1}^2 - m_{{\tilde t}_2}^2}\, \frac{9}{8} \, c_{2t} \Bigg\} \nonumber \\
 &+C_A\, C_F^2\, \Bigg\{ \frac{m_{\tilde g}\, m_{t}}{m_{{\tilde t}_1}^2 - m_{{\tilde t}_2}^2}\, \frac{9}{8} \, c_{2t} - \frac{3}{32}\, c_{2t}\, s_{2t} \Bigg\} \nonumber \\
& +C_F^2\, T_f\, \Bigg\{ n_q\, \Bigg[ 
\frac{m_{\tilde g}\, m_{t}}{ m_{{\tilde t}_1}^2 - m_{{\tilde t}_2}^2 } \, c_{2t}\left( 3  -\frac{9}{2} \, \zeta_3 \right)
+ c_{2t}\, s_{2t} \, \left( \frac{3}{8}\, \zeta_3 -\frac{1}{4} \right) \Bigg] \nonumber \\
&+ n_t\, \Bigg[ 
\frac{m_{\tilde g}\, m_{t}}{ m_{{\tilde t}_1}^2 - m_{{\tilde t}_2}^2 } \, c_{2t}\left( 3  -\frac{9}{2} \, \zeta_3 \right)
+ c_{2t}\, s_{2t} \, \left( \frac{3}{8}\, \zeta_3 -\frac{1}{4} \right) \Bigg] \nonumber \\ 
& +C_A\, C_F\, T_f\, \Bigg\{ n_q\, \Bigg[
\frac{m_{\tilde g}\, m_{t}}{ m_{{\tilde t}_1}^2 - m_{{\tilde t}_2}^2 } \, c_{2t} \, \left(
 \frac{9}{2} \, \zeta_3 -\frac{3}{8} \right)
+ c_{2t}\, s_{2t} \, \left( \frac{1}{32} - \frac{3}{8}\, \zeta_3\right) \Bigg] \nonumber \\
&+ n_t\, \Bigg[
\frac{m_{\tilde g}\, m_{t}}{ m_{{\tilde t}_1}^2 - m_{{\tilde t}_2}^2 } \, c_{2t} \, \left(
 \frac{9}{2} \, \zeta_3 -\frac{3}{8} \right)
+ c_{2t}\, s_{2t} \, \left( \frac{1}{32} - \frac{3}{8}\, \zeta_3\right) \Bigg]
\Bigg\}\,.% \nonumber 
\label{eq::theta3}
\end{align}

At that point a brief comment on degenerate top squarks is in order. 
In the expressions for $\gamma_{m_{\tilde{t}_i}}$ the limit
$m_{{\tilde t}_2}\to m_{{\tilde t}_1}$ can be taken
naively. Furthermore one has to nullify the mixing angle.
The quantity $\gamma_{\theta_t}$ is not defined in the  mass-degenerate
case which is reflected by the fact that the limit $m_{{\tilde t}_2}\to
m_{{\tilde t}_1}$ does not exist in 
Eqs.~(\ref{eq::gamma0}),~(\ref{eq::theta2}) and~(\ref{eq::theta3}).

In order to compare with the results in the literature we have to transform
our results to the anomalous dimensions for the quantities $M_{\tilde{Q}}$,
$M_{\tilde{U}}$ and $A_t$ as given in Eq.~(\ref{eq::Mtil}). This is
conveniently achieved with the help of Eq.~(\ref{eq::Mtildiag}) which
is differentiated w.r.t. $\mu^2$. The resulting equations are then solved for
the $\gamma_{M_{\tilde{Q}}}$, $\gamma_{M_{\tilde{U}}}$ and
$\gamma_{A_t}$. We have compared the resulting one-, two- and three-loop
expressions with the results in the
literature~\cite{Barger:1993gh,Jack:1994kd,Jack:2004ch} and found 
complete agreement. Note that the method used in Ref.~\cite{Jack:2004ch} is
based on a relation of the anomalous dimensions to an all-order expression
in the so-called NSVZ scheme~\cite{Novikov:1983uc} 
whereas in this work a diagrammatic approach has been used
to evaluate the three-loop corrections.
We refrain from providing explicit results for
$\gamma_{M_{\tilde{Q}}}$ and $\gamma_{M_{\tilde{U}}}$ which, however,
can be found in the {\tt Mathematica} file~\cite{progdata}.
Note that we have $\gamma_{M_{\tilde{Q}}}=\gamma_{M_{\tilde{U}}}$ which
is expected since electroweak effects are neglected~\cite{Avdeev:1997vx}. 
This serves as a welcome check for our calculation.
The result for $\gamma_{A_t}$ is proportional to the gluino mass and is
thus quite compact. Up to three-loop order it is given by
\begin{align}
 \frac{\mu^2}{A_t} \, \frac{d}{d\mu^2} \, A_t &= \gamma_{A_t} =
 \frac{m_{\tilde g}}{A_t} \, \Bigg\{ \frac{\alpha_s}{\pi}\, C_F
 + \left( \frac{\alpha_s}{\pi} \right)^2 \, \Bigg[ \frac{3}{2} \, C_A\, C_F - C_F^2 - C_F\, \left( n_q + n_t \right)\, T_f \Bigg] \nonumber \\
 &+ \left( \frac{\alpha_s}{\pi} \right)^3 \, \Bigg[ \frac{9}{8}\, C_A^2\, C_F - \frac{9}{8}\, C_A\, C_F^2
 + \frac{3}{2} \, C_F^3 - \frac{3}{4} \, C_F\, \left( n_q + n_t \right)^2\, T_f^2 \nonumber \\
& +C_A\, C_F\, \left( n_q + n_t \right)\, T_f\, \left( \frac{3}{8} - \frac{9}{2} \, \zeta_3 \right) 
 +C_F^2\, \left( n_q + n_t \right)\, T_f\, \left( -3 +\frac{9}{2} \, \zeta_3 \right)
\Bigg]
\Bigg\}\,.
\end{align}

For completeness let us also provide the result for mass-degenerate
squarks which is given by
\begin{align}
  m_{\tilde q}^2 \gamma_{m_{\tilde{q}}}^{(0)}
  &= C_F \, m_{\tilde{g}}^2\,,% \nonumber 
\\
%\end{align}
%\begin{align}
  m_{\tilde q}^2 \gamma_{m_{\tilde{q}}}^{(1)}
  &= 
C_A\, C_F\, \Bigg\{ \frac{3}{4}\, m_{\epsilon}^2 + \frac{11}{4}\, m_{\tilde g}^2 \Bigg\}
- C_F^2\, \frac{3}{2}  \, m_{\tilde g}^2 \nonumber \\
& - C_F\, \, T_f \, \Bigg\{ n_q \, \Bigg[ \frac{1}{2} \, m_{\epsilon}^2 + \frac{3}{2}\, m_{\tilde g}^2 + m_{\tilde{q}}^2 \Bigg]
+ n_t \,  \Bigg[ \frac{1}{2} \, m_{\epsilon}^2 + \frac{3}{2} \, m_{\tilde g}^2 + \frac{1}{2} \, m_{{\tilde t}_1}^2
+ \frac{1}{2}\, m_{{\tilde t}_2}^2 - m_{t}^2 \Bigg] \Bigg\}
\,,%\nonumber 
\\
%\end{align}
%\begin{align}
  m_{\tilde q}^2 \gamma_{m_{\tilde{q}}}^{(2)}
  &= 
C_F^3\, 3\, m_{\tilde g}^2
- C_A\, C_F^2\, \Bigg\{ \frac{9}{16} \, m_{\epsilon}^2 + \frac{21}{8} \, m_{\tilde g}^2 \Bigg\}
+C_A^2\, C_F\, \Bigg\{ \frac{45}{32}\, m_{\epsilon}^2 +\frac{15}{4}\, m_{\tilde g}^2 \Bigg\} \nonumber \\
& + C_F\, T_f^2 \, \Bigg\{ n_q^2 \, \Bigg[ \frac{3}{8} \, m_{\epsilon}^2 - \frac{3}{2} \, m_{\tilde g}^2 + \frac{3}{4} \, m_{\tilde{q}}^2 \Bigg] \nonumber \\
&+ n_q\, n_t\, \Bigg[ \frac{3}{4} \, m_{\epsilon}^2
-3\, m_{\tilde g}^2 + \frac{3}{4}\, m_{\tilde{q}}^2 + \frac{3}{8} \, m_{{\tilde t}_1}^2 + \frac{3}{8} \, m_{{\tilde t}_2}^2 
-\frac{3}{4} \, m_{t}^2 \Bigg] \notag\\
 &+ n_t^2\, \Bigg[ \frac{3}{8} \, m_{\epsilon}^2 - \frac{3}{2} \, m_{\tilde g}^2 + \frac{3}{8} \, m_{{\tilde t}_1}^2 + \frac{3}{8} \, m_{{\tilde t}_2}^2 
-\frac{3}{4}\, m_{t}^2 \Bigg] \Bigg\} \nonumber \\
& -C_A\, C_F\, T_f\, \Bigg\{ n_q\, \Bigg[ \frac{3}{2} \, m_{\epsilon}^2 
+\frac{15}{8} \, m_{\tilde{q}}^2 + 9\, m_{\tilde g}^2\, \zeta_3 \Bigg] \nonumber \\
& +n_t\, \Bigg[ \frac{3}{2}\, m_{\epsilon}^2 + \frac{15}{16} \, m_{{\tilde t}_1}^2 
+\frac{15}{16}\, m_{{\tilde t}_2}^2 - \frac{15}{8}\, m_{t}^2 + 9\, m_{\tilde g}^2\, \zeta_3 \Bigg] \Bigg\} \nonumber \\
&+C_F^2\, T_f\, \Bigg\{ n_q\, \Bigg[ \frac{3}{8}\, m_{\epsilon}^2 
-\frac{27}{4}\, m_{\tilde g}^2 + \frac{3}{4}\, m_{\tilde{q}}^2 + 9\, m_{\tilde g}^2\, \zeta_3 \Bigg] \nonumber \\
&+n_t\, \Bigg[ \frac{3}{8}\, m_{\epsilon}^2
-\frac{27}{4}\, m_{\tilde g}^2 + \frac{3}{8}\, m_{{\tilde t}_1}^2 + \frac{3}{8} \, m_{{\tilde t}_2}^2 - \frac{3}{4}\, m_{t}^2
+9\, m_{\tilde g}^2\, \zeta_3 \Bigg] \Bigg\}\,.% \nonumber 
\end{align}
One observes that all terms which do not involve $n_t$ can be obtained
from $\gamma_{m_{\tilde{t}_1}}$ by setting
$m_{\tilde{t}_2}=m_{\tilde{t}_1}$, $m_t=0$ and $\theta_t=0$. 

When applying the anomalous dimensions derived in this paper one has
to consider the combined set of differential equations of all $\drbar$
parameters appearing on the r.h.s. of the above results. This concerns
in particular the unphysical $\epsilon$-scalar mass which means
that although $m_\epsilon$ is set to zero at one scale it is
different from zero once this scale is changed.
A way out from this situation is to renormalize the $\epsilon$ scalar
mass on-shell. We have computed the resulting anomalous dimensions and
provide the results in Ref.~\cite{progdata}. Alternatively one could
shift the squark masses by a finite term which is chosen such that the
$\epsilon$ scalar decouples from the system of differential equations.
The resulting renormalization scheme is called $\drbarprime$ scheme
and has been suggested in Ref.~\cite{Jack:1994rk}. In our
approximation the finite shift is needed up to two loops which is given
by~\cite{Jack:1994rk,Martin:2001vx}
\begin{align}
  m^2_{\tilde{f}} \rightarrow  m^2_{\tilde{f}} - \frac{\alpha_s}{\pi} \,
  \frac{1}{2} \, C_F \, m^2_{\epsilon} 
  + \left( \frac{\alpha_s}{\pi} \right)^2 \, C_F \, m^2_{\epsilon} \, \left(
    \frac{1}{4} \, T_f \, \left( n_q + n_t \right) + \frac{1}{4} \, C_F - \frac{3}{8} \, C_A
  \right)
  \,,
\end{align}
where $f=t$ or $f=q$.\footnote{Of course, $T_f$ is not altered.}
We have checked that after inserting this shift in
$\gamma_{m_{\tilde{t}_1}}$, $\gamma_{m_{\tilde{t}_2}}$ and
$\gamma_{m_{\tilde{q}}}$ the parameter
$m_\epsilon$ drops out from the resulting anomalous dimension.
Again we refrain from listing explicit results, however, provide the
analytic expressions in~\cite{progdata}.

All results presented above can be found in {\tt Mathematica} format
on the webpage~\cite{progdata}. In addition we provide the
results for the anomalous dimensions $\gamma_{M_{\tilde{Q}}}$,
$\gamma_{M_{\tilde{U}}}$ and $\gamma_{A_t}$ and the renormalization constants 
for the squark masses and the mixing angle in the top squark system.
The {\tt Mathematica} file contains furthermore the result for
$\gamma_{m_{\tilde{t}_1}}$, $\gamma_{m_{\tilde{t}_2}}$ and
$\gamma_{m_{\tilde{q}}}$ for on-shell $\epsilon$ scalar
masses and in the $\drbarprime$ scheme.

Let us finally perform a simplified analysis in order to
exemplify the numerical impact of the three-loop corrections. In our 
example we fix the following values of the $\drbarprime$ 
parameters at the scale $\mu=\mu_G=10^{16}$~GeV
\begin{align}
  &m_{\tilde{t}_1}=400~\mbox{GeV}\,,\quad
  m_t=67~\mbox{GeV}\,,\quad 
  \theta_t=0.1\,,\quad 
  \alpha_s = 0.0425\,, \nonumber\\
  &m_{\tilde{t}_2}=m_{\tilde{g}}=m_{\tilde{q}}=600~\mbox{GeV}\,,
  \label{eq::num}
\end{align}
and use the anomalous dimensions obtained in this paper and in
Ref.~\cite{Harlander:2009mn} to compute the corresponding values for
$\mu=M_Z$. Since our aim is to study the numerical importance of the 
three-loop anomalous dimensions in the squark sector we neglect all threshold
effects. Furthermore, we use for the running of $\alpha_s$, $m_{\tilde{g}}$
and $m_t$ always the three-loop approximation whereas in the case of the
squark masses and $\theta_t$ the loop-order is varied from one to three.

The values of $m_t=m_t(\mu_G)$ and $\alpha_s=\alpha_s(\mu_G)$ in
Eq.~(\ref{eq::num}) are chosen such
that three-loop running leads to $m_t(M_Z)=170$~GeV and
$\alpha_s(M_Z)=0.118$. 
The results for $m_{\tilde{t}_1}$, $m_{\tilde{t}_2}$, $\theta_t$ and
$m_{\tilde{q}}$ at the scale $\mu=M_Z$ can be found in Tab.~\ref{tab::run}.

\begin{table}
  \begin{center}
    \begin{tabular}{c|ccc}
      & 1 loop & 2 loops & 3 loops \\
      \hline
      $m_{\tilde{t}_1}$ (GeV) &1425 &1416& 1378\\
      $m_{\tilde{t}_2}$ (GeV) &1677 &1670& 1632\\
      $\theta_t$              &0.658&0.659&0.656\\
      $m_{\tilde{q}}$ (GeV)   &1580 &1573& 1535\\
    \end{tabular}
    \caption{\label{tab::run}Numerical values for the $\drbarprime$ 
      parameters for $\mu=M_Z$ using the numbers in
      Eq.~(\ref{eq::num}) as input and solving the system of
      differential equations with one-, two- or three-loop anomalous dimensions
      in the squark sector.}
  \end{center}
\end{table}

We observe a small change in the mixing angle by about 0.4\%.  As far as the
squark masses are concerned one observes a moderate shift of a few GeV when
going from one to two loops. After switching on the three-loop terms, however,
the squark masses are decreased by about 40~GeV which is approximately an
order of magnitude larger than the two-loop corrections.  Nevertheless it
corresponds to a shift in the masses of about 3\% which is a reasonable amount
for a three-loop SUSY QCD term.  Our observation coincides with the findings
of Ref.~\cite{Jack:2004ch} where also relatively large three-loop corrections for
the squarks have been identified.

%- }}}
%- {{{ Conclusions

\section{Conclusions}\label{sec::conclusions}

In this paper the renormalization constants for the squarks and the
corresponding mixing angle have been computed to three-loop
order within supersymmetric QCD. Thus, all anomalous dimensions of
the physical parameters are now available to order $\alpha_s^3$ and can
thus be used to relate their mass values at the GUT and electroweak
or TeV scale. 

Our calculation has been performed using dimensional reduction for the
regularization of the divergent loop integrals which is realized with
the help of massive $\epsilon$ scalars. As far as the renormalization
of the $\epsilon$ scalar mass is concerned we have evaluated our
results for three different schemes: $\drbar$, $\drbarprime$ and on-shell.
Our results agree with Ref.~\cite{Jack:2004ch} which supports the
consistency of DRED with SUSY QCD since in Ref.~\cite{Jack:2004ch} the
results have been obtained without a diagrammatic calculation.

A simplified numerical analysis shows that the three-loop corrections to the
squark masses are numerically important (see also~\cite{Jack:2004ch}) and thus
should be included in the spectrum generators which incorporate the running
from the GUT to the electroweak scale.

All renormalization constants and anomalous dimensions computed in this
paper can be downloaded from the URL~\cite{progdata} 
in {\tt Mathematica} format.

%- }}}
%- {{{ Ackn.
 
\section*{Acknowledgements}

This work was supported by the DFG through the SFB/TR~9 ``Computational
Particle Physics'' and the Graduiertenkolleg ``Elementarteilchenphysik bei
h\"ochster Energie und h\"ochster Pr\"azision''.

%- }}}

%\begin{appendix}
%\end{appendix}

%- }}} body:
%- {{{ bibliography:

%- }}}

\end{document}